\documentstyle[11pt,epsf,newpasp,twoside]{article}
\newcommand{\basel}{BaSeL~}
\newcommand{\teff}{$T_{\mathrm{eff}}$}
\newcommand{\logg}{$\log g$}

\newcommand{\feh}{$[Fe/H]$}

\begin{document}

\title{Photometric properties of theoretical spectral libraries for
GAIA photometry}

\author{T. Lejeune}
\affil{Observat\'orio Astron\'omico da Universidade de Coimbra, Portugal}
\begin{abstract}
Photometric properties of various spectral libraries of synthetic
stellar spectra derived from the ATLAS 9 (Castelli et al. 1997), the
PHOENIX NextGen (Hauschildt et al. 1998, 1999), the NMARCS (Bessell et
al. 1998), and the \basel (Lejeune et al. 1998, Westera et al. 2002)
models are discussed in view of their application to the definition of
the GAIA photometric system. Comparisons with empirical UBVRI
calibrations show significant discrepancies at low temperatures between
the different models, and with respect to the \basel 3.1 and the
Alonso et al. (1998, 1999) empirical calibrations.
\end{abstract}

\section{Introduction}
The GAIA Galactic survey will observe more than 1 billion of stars in
our Galaxy, and will obtain photometry in 11 medium-band and 5
large-band filters with the challenging goal to determine the
fundamental stellar parameters (\teff, \logg, \feh) across a very wide
range of stellar types.  In order to test the capabilities of the
photometric system and the performances of the classification
algorithms, various stellar libraries of synthetic and/or empirical
spectra are required.  In this study, I compare the photometric
properties of the most widely used grids of synthetic spectra (ATLAS,
NMARCS, PHOENIX, BaSeL) for cool stars in view of their application
to GAIA photometry.
 
\section{Grids of synthetic stellar spectra} 

The coverage in stellar parameters for each of the grids of models
used in this comparative study is given in Table \ref{tab:data}.  The
BaSeL 2.2 hybrid spectral library was constructed from empirical
($[Fe/H]=0$) and semi-empirical ($[Fe/H] \neq 0$) color-temperature
relations (see Lejeune et al.  1997, 1998, for details), while the new
BaSeL 3.1 models are based on purely empirical calibrations, defined
for the metallicity range $-2.0 \leq [Fe/H] \leq 0.0$ from a large
collection of globular cluster UBVRIJHKL photometric data (see Westera
et al. 2002 for details). Hence, the BaSeL 3.1 calibrations provide
the only existing set of empirical metallicity-dependent
\teff-UBVRIJHKL transformations for $-2.0 \leq [Fe/H] \leq 0.0$ over a
large temperature range, from 2000 K to 50000 K, and are used as
reference in this study. Synthetic spectra from the ATLAS 9 atmosphere
models have been computed by Castelli et al. (1997) with no
overshooting parameter.

\begin{table*}[hbt]		
\begin{flushleft} 	
\caption[]{Parameter coverage of the different grids of models}
\label{tab:data}
\smallskip
\begin{center}	
\small
\begin{tabular}{ccccccc}	
\tableline\noalign{\smallskip} 
Models & Note & \teff & \logg & \feh & $\lambda$ (nm) \\
\noalign{\smallskip}		
\tableline\noalign{\smallskip} 
\basel 2.2$^{(1)}$  &
hybrid lib. &
$2000$ - $50000$ K &
$-1.0$ - $5.5$ &
$-5.0$ - $1.0$ &
$9.1$ - $1.6\, 10^5$ \\
\noalign{\smallskip}	
\tableline\noalign{\smallskip} 
\basel 3.1$^{(2)}$ &
hybrid lib. &
$2000$ - $50000$ K &
$-1.0$ - $5.5$ &
$-2.0$ - $0.5$ &
$9.1$ - $1.6\, 10^5$ \\
\noalign{\smallskip}	
\tableline\noalign{\smallskip} 
ATLAS 9$^{(3)}$  &
no overs. &
$3500$ - $50000$ K &
$0.0$ - $5.0$ &
$-2.5$ - $0.5$ &
$9.1$ - $1.6\, 10^5$  \\
\noalign{\smallskip}	
\tableline\noalign{\smallskip} 
PHOENIX &
giants$^{(4)}$ &
$2000$ - $7000$ K &
$-0.7$ - $0.0$ &
$-0.7$ - $0.0$ &
$10$ - $10^6$ \\
\noalign{\smallskip}
NextGen &
dwarfs$^{(5)}$ &
$1000$ - $7000$ K &
$3.5$ - $6.0$ &
$-4.0$ - $0.0$ &
$10$ - $ 10^6$ \\
\noalign{\smallskip}	
\tableline\noalign{\smallskip} 
NMARCS$^{(6)}$ &
giants &
$3600$ - $4750$ K &
$-0.5$ - $3.5$ &
$-0.6$ - $0.6$ &
BVRIJHKL   \\
\noalign{\smallskip}
-- &
dwarfs &
$2600$ - $4000$  K &
$4.5$ - $5.0$ &
$-2.0$ - $0.3$ &
BVRIJHKL  \\
\noalign{\smallskip}	
\tableline	
\end{tabular}	
\end{center}	
\smallskip
\scriptsize 	
$^{(1)}$ Lejeune et al. 1997, 1998; \\	
$^{(2)}$ Westera et al. 2002; \\
$^{(3)}$ Castelli et al. 1997; \\
$^{(4)}$ Hauschildt et al. 1999a; $^{(5)}$ Hauschildt et al. 1999b; \\
$^{(6)}$ Bessell et al. 1998.
\end{flushleft} 
\end{table*}

\section{Comparisons in the two-color diagrams}
\begin{figure*}[!h]
	\centering 
	\leavevmode \centerline{\epsfxsize=.65\textwidth
	\epsfbox{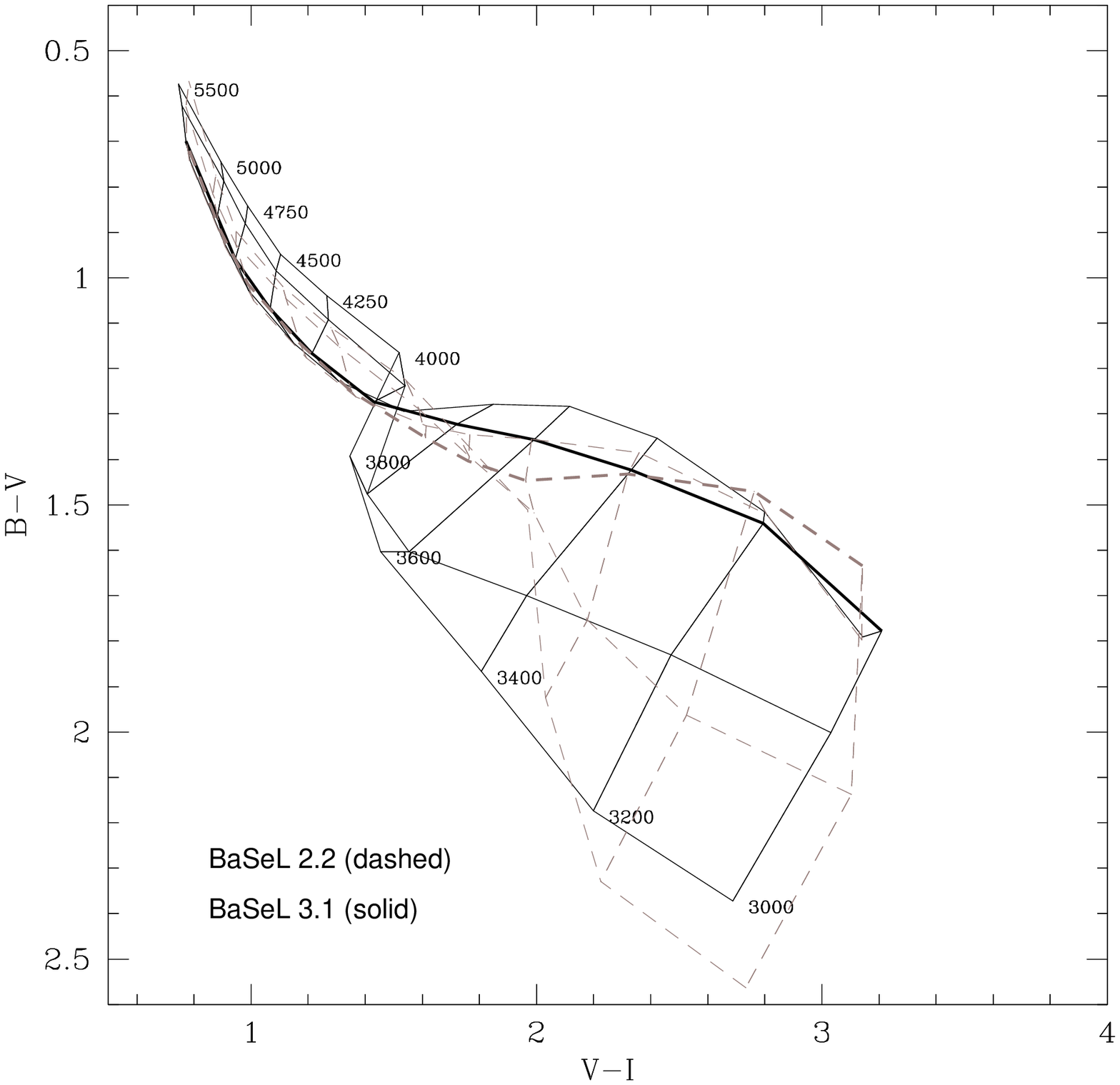}} \centering
	\leavevmode \centerline{\epsfxsize=.65\textwidth
	\epsfbox{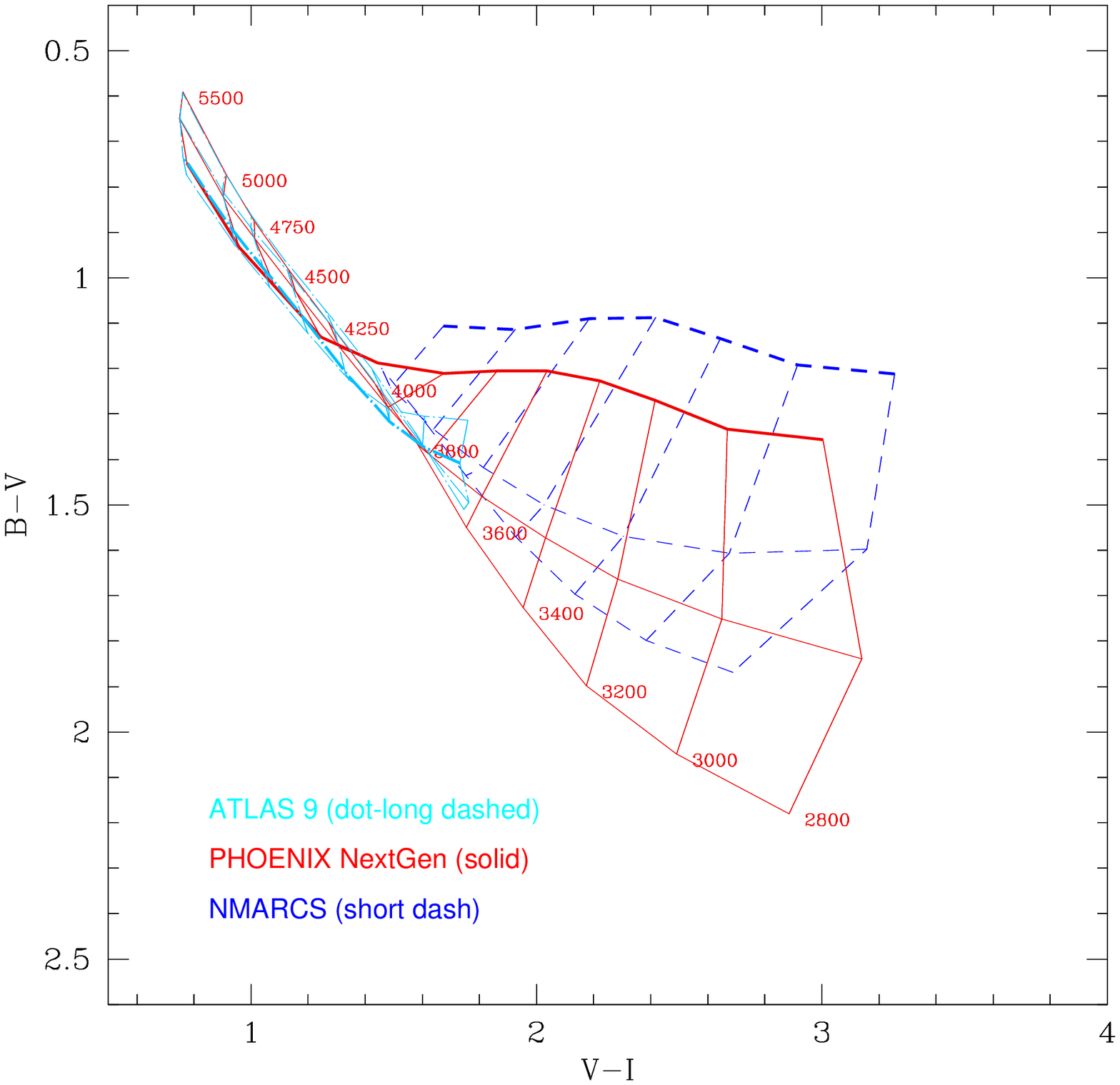}}
	\caption{A comparison of the different grids of models with
	\logg =1 in the (B-V)/(V-I) two-color diagram. Top panel: the
	{\basel} 2.2 and 3.1 model colors. Bottom panel: the ATLAS 9
	(Castelli et al.), the PHOENIX and the NMARCS models. The
	lines connect the models with the same metallicity (\feh\
	$=-0.6, -0,3$ and $0.0$ (thick line), and the models with the
	same \teff\ as indicated in each panel. The models show very
	large differences in the M giant regime (see text).}
\label{f:two_color_diag}
\end{figure*}

In Fig. \ref{f:two_color_diag}, model colors in the range $2500 K
\leq$ \teff $\leq 5500 K$, computed from the different grids of
synthetic spectra given in Table \ref{tab:data}, are compared in the
\linebreak (B-V)/(V-I) color-color diagram for the same value of the
surface gravity ($\log g=1$) with {\feh} $=-0.6, -0.3$ and $0.0$. From
the figure, it is clear that large differences exist between the
different grids, specially in the low temperature regime (\teff\ $<
4000 K$). Above 4000 K, the NMARCS models provide the best overall
agreement with the \basel 3.1 empirical calibrations, both in B-V and
V-I, in particular for the solar metallicity. In contrast, the PHOENIX
models appear too blue in B-V, while they are in good agreement with
the \basel 3.1 V-I colors. The B-V model colors from the ATLAS 9 grid
are systematically redder below 4500 K. Below 4000 K, very large
differences of several tenth of magnitude exist between all the model
colors.

Similar comparisons (Lejeune et al., in prep.) for the dwarf models
show that large deviations between the models and the \basel 3.1
empirical calibrations also exist, and are maximum below 4000 K. For
the M dwarfs, model colors computed from the PHOENIX models provide
the best agreement with the \basel 3.1 empirical calibrations.

\section{Temperature-color calibrations}
We also compared the theoretical temperature-color calibrations,
\teff-(U-B) and \teff-(B-V), predicted by each grid of models in the
range $3000 K \leq$ \teff\ $\leq 6000 K$, with the temperature scales
adopted in \basel 3.1 and the empirical calibrations of Alonso et al
(1998, 1999) derived from the IRFM method. Our results (Lejeune et
al., in prep) show that, for \teff\ $> 4000 K$, the theoretical scales
for all the models agree well with the BaSeL 3.1 empirical relations
for \feh\ $= 0.0$ and $-2.0$. Surprisingly, below 4000 K the
theoretical calibrations from ATLAS 9 model atmospheres are found in
very good agreement with the empirical values. The \teff-(B-V)
relation from the NMARCS model colors for dwarfs deviates
significantly (more than 0.5 mag at \teff\ $= 3000 K$) from the \basel
3.1 empirical calibrations, while the NextGen model calibrations are
closer. For the giants, the NMARCS models provide in general a better
match to the empirical relations than the NextGen models. Comparisons
of the \basel 3.1 calibrations with the metallicity-dependent
temperature-color empirical relations of Alonso et al. in the
temperature range 7000 K -- 4000 K show a good agreement (less than
0.05 mag) for the dwarf sequences, but a more significant discrepancy
($>$ 0.1 mag in average) for the giant sequence with \feh\ $= -2.0$.

\section{Conclusions}
UBVRI colors of the most widely used theoretical spectral libraries
(ATLAS 9, PHOENIX, NMARCS, and \basel) in the temperature range 7000 K
-- 3000 K have been compared with empirical calibrations available for
$-2.0 \leq [Fe/H] \leq +0.5$. We found that the ATLAS 9 model colors
from Castelli et al. (1997) agree well or very well with the empirical
color-temperature relations at all metallicities and over the range of
effective temperatures between 6000 K and 3500 K. The most significant
deviations are found for the \teff-(U-B) relation at solar
metallicity.  Below 4000 K, important deviations are found for the UBV
colors between both the PHOENIX NextGen and the NMARCS model
predictions with respect to the \basel 3.1 empirical relations,
although some uncertainties still exist in the empirical data at these
low temperatures.  In a general way, the deviations appear to be
less important with the PHOENIX NextGen model colors for dwarfs, and
with the NMARCS models for giants.

These differences have to be accounted for, or maybe reduced with some
spectral corrections methods, in the selection of models in order to
construct a spectral library well suitable to the definition of
the GAIA photometric system.

\acknowledgments This work was  supported by the ``Funda\c{c}\~ao para
a     Ci\^encia    e    a     Tecnologia''    through     the    grant
PRAXIS-XXI/BPD/22061/99),        and       by        the       project
``PESO/P/PRO/15128/1999''.

%%%%%%%%%%%%%%%%%%%%%%%%%%%%%%%%%%%%%%%%%%%%%%%%%%%%%%%%%%%%%%%%%%%%%

%%%%%%%%%%%%%%%%%%%%%%%%%%%%%%%%%%%%%%%%%%%%%%%%%%%%%%%%%%%%%%%%%%%%%

\end{document}